# Considerations of Earth' climate sensitivity based on peculiarities of planetary heat capacity using system identification method: Runaway greenhouse effect scenario is still possible

Alexei V Karnaukhov[1], Sergei F Lyuksyutov[2], Artem V Aliakin[3],

Mikhail E Prokhorov[3], and Sergei I Blinnikov[4]

[1]*Institute of Cell Biophysics, Russian Academy of Sciences Pushchino 142290, Russia*

[2]*The University of Akron, Physics Department, Akron OH 44325, United States of America*

[3]*Lomonosov Moscow State University, Physics Department, 119991 Moscow, Russia*

[4]*Lomonosov Moscow State University, Sternberg Astronomical Institute, 119991 Moscow, Russia*

[5]*Kurchatov Complex for Theoretical and Experimental Physics, 117218 Moscow, Russia*

E-mail: sfl@uakron.edu (Sergei F Lyuksyutov)

## Abstract

System identification method (SIM) was used to evaluate the Earth's equilibrium climate sensitivity (ECS). According to our simulations, the ECS was found to be between 2.0°C and 7.0°C. Analysis of the changes in heat inventory of oceans, atmosphere, land, and cryosphere was based on the experimental data of IPCC6.

The equation derived for Earth's global surface temperature (GST) shows that the sum of the dimensionless feedback coefficients from water vapor, methane, and Earth's albedo could be less than 1. However, due to the positive feedback from carbon dioxide (the combined greenhouse catastrophe) and the revised ECS estimate based on an increase in GST (leading to an increase in ECS), the probability of the runaway greenhouse effect increases significantly. It is still less than the critical number when not considering the feedback associated with carbon dioxide, water vapor, and methane buildup in Earth's atmosphere.

The analysis considers the thermal dynamics of the oceans and other factors, including the exponential growth of the Earth's global temperature based on IPCC6 data.

*Keywords:* Earth' climate sensitivity, Planetary heat capacity, System identification method

## 1. Introduction

Across our planet, the critical connections between the ocean and the top of the atmosphere (TOA) have been disrupted. The stability that we and all our life on the planet rely upon is being lost. It is highly likely that the global temperature limit (~3 °C+) could be exceeded in the next twenty years. Even a 2°C increase in global warming would be catastrophic, creating unpredictable heatwaves, droughts, extreme precipitation, and wildfires. It has been conservatively estimated that since the First Industrial Revolution, the Earth has warmed between 1.5 °C and 1.6 °C.

Earth's energy budget encompasses the major energy flows relevant to the climate system, as presented in Figure 1. The top-of-atmosphere (TOA) energy budget is determined by the incoming short-wavelength solar radiation and the outgoing long-wave radiation. However, anthropogenic forcing has created an imbalance in the global mean TOA radiation budget, which is an important metric for the rate of global climate change. It is also a major driver of the global water cycle, atmosphere and ocean dynamics, as well as various surface processes [1].

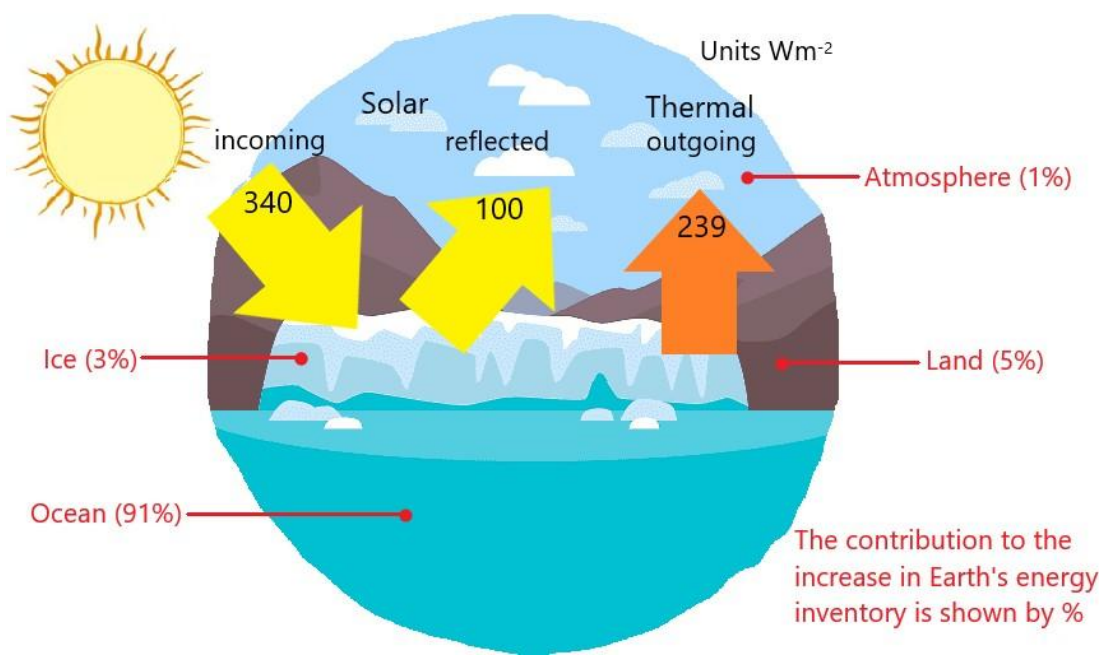

**Figure 1.** The global energy budget is based on the IPCC6 data. The intensity units are in W m$^{-2}$

The energy budget for land and ocean is still a subject of considerable uncertainty. Numerous evaluations of the budget have been reported by Wild et al. using CMIP5 climate models [1], based on direct observations from the surface and space [2], the analysis of multi-century pre-industrial control simulations by Palmer and McNeall [3], by Roberts et al. [4], and Earth's energy imbalance (EEI) as a fundamental metric instead of global surface temperature by von Schuckmann et al. in [5]. Estimations of long-term upper-ocean warming by Durack et al. [6], integrated upper-ocean heat content anomalies in [7] by Lyman and Johnson, and the assessment of ocean heat content (OHC) have been accomplished, indicating that OHC increased steadily by Cheng et al. [8]. The "efficacy" (global temperature response per unit forcing relative to the response to CO2 forcing) varies substantially among climate forcing agents, with climate forcing $\sim 0.8$ W m$^{-2}$ for the period 1750-2000 making CH4 apparently the most anthropogenic climate forcing other than CO2 by Hansen et al. [9]. The change of TOA energy flux as a function of effective radiative forcing, the temperature change, and the feedback parameter introduced in [9] and was used for the equilibrium climate sensitivity (ECS) estimate in this work. It was suggested that the ECS estimates are very uncertain and highly likely within 1.5–4.5 °C by Zelinka et al. [10]. Palmer et al. [11] show that global mean surface temperature responds relatively quickly to changes in emissions, leading to a negative trend in post-2100, although the temperature remains substantially elevated compared to the present day up to 2300. In contrast, EEI remains positive and results in ongoing sea-level rise from global thermal expansion. Williams et al. [12] report that providing tighter constraints on how much carbon may be emitted based on the transient climate response to cumulative carbon emissions requires providing tighter bounds for estimates of the physical climate feedback, particularly from clouds, as well as to a lesser extent for other contributions from the rate of ocean heat uptake, the terrestrial and ocean cycling of carbon. Pfister and Stocker [13] suggest that reduced-complexity models remain useful tools for future climate change projections but should employ a range of climate sensitivity tunings to account for the uncertainty in both the long-term warming and the realized warming fraction. Huusko et al. [14] show that, over the 20th century, there is a weak correlation between total forcing and ECS, contributing to, but not determining, the model agreement with observed warming. The ECS and aerosol forcing in the models are not correlated. Tokarska et al. [15] find that ocean warming simulations are consistent with greenhouse gas increases from observations. Other models show the feedback during the historical period may differ from the feedback at $CO_2$ doubling and from those at true equilibrium. Rogelj et al. [16] suggest that, to stabilize global-mean temperature at levels of 2 °C or lower, strong reductions of greenhouse gas emissions to stay within the allowed carbon budget seem therefore unavoidable over the 21st century. Sherwood et al. [17] calculate probability distributions of the committed warming that would ensue if all anthropogenic emissions were stopped immediately, or at successive future times. This analysis reveals a wide range of possible outcomes, including no further warming, but also a 15% chance of overshooting the 1.5 °C target, and 1% – 2% chance for 2 °C, even if all emissions had stopped in 2020. If emissions merely stabilize in 2020 and stop in 2040, these probabilities increase to 90% and 17%. The uncertainty arises mainly from that of present forcing by aerosols. Rather than there being a fixed date by which emissions must stop, the probability of reaching either target, which is already below 100%, gradually diminishes with delays in eliminating emissions, by 3%–4% per year for 1.5 °C. Zhou et al. [18] show that, after the pattern effect is accounted for, the best-estimate value of committed global warming at present-day forcing rises from 1.31 °C (0.99–2.33 °C, 5th–95th percentile) to over 2 °C, and committed warming in 2100 with constant long-lived forcing increases from 1.32 °C (0.94–2.03 °C) to over 1.5 °C, although the magnitude is sensitive to the sea surface temperature dataset. Further constraints on the pattern effect are needed to reduce climate projection uncertainty. Dassler [19] investigates potential biases between equilibrium climate sensitivity inferred from warming over the historical period (ECShist) and the climate system's true ECS (ECStrue). The net effect of the pattern effect can produce an estimate of ECShist as much as 0.5 °C below ECStrue. Dessler and Forster [20] see no evidence to support low ECS (values less than 2 °C) suggested by other analyses. They estimate that ECS is likely 2.4–4.6 °C (17–83% confidence interval), with a mode and median value of 2.9 and 3.3 °C, respectively. Dassler et al. [21] find that framing energy balance in terms of 500 hPa tropical temperature better describes the planet's energy balance.

The Earth's climate stability is the "to be or not to be" question for humankind. Studies on positive feedback between rising global temperatures and the amount of water vapor in the Earth's atmosphere have predicted that the runaway greenhouse effect may inevitably happen on our planet in only a few hundred million years from now [22–31]. The runaway greenhouse effect scenario caused by anthropogenic forcing and an uncontrollable increase in Earth's TOA caused by the positive feedback is named the "moist greenhouse catastrophe" (an increase in global temperature results in an increase in water vapor concentration).

It is speculated that our Sun may increase its luminosity by 10% every billion years [32, 33]. Currently, the Sun's luminosity is still insufficient for the "moist greenhouse catastrophe" scenario. However, the additional positive feedback caused by raising the water vapor level in Earth's atmosphere could significantly increase the ECS. This opens a window for a possible runaway greenhouse scenario because of that feedback, even if not related to water vapor accumulation in the atmosphere.

The first feedback could be ultimately related to the reserves for methane hydrates in the permafrost zone and on the continental shelf. Methane hydrates are compounds of methane and water that are only stable at high pressure and low temperature. An increase in global temperature can lead to the decomposition of accumulated reserves of methane hydrates and could potentially release additional amounts of methane into the atmosphere. Since methane is a greenhouse gas, this process could lead to an additional increase in the Earth's surface temperature. The analysis of positive feedback increases in global temperature suggests (i) decomposition of methane oxides, (ii) the increase in methane concentration in the atmosphere, (iii) conversion of methane to $CO_2$, (iv) the increase in global temperature [34–37]. Based on the analysis, the scenario of a catastrophic uncontrollable increase in temperature on Earth due to the decomposition of methane hydrates seems unlikely today. A "clathrate gun" alone will most likely not fire.

Methane hydrates are not the only reservoir of greenhouse gases on our planet. With rising global temperatures, they could lose stability and lead to the release of significant amounts of greenhouse gases into the atmosphere. For example, in the permafrost zone, significant amounts of organic matter are frozen. As global temperatures rise, this organic matter will thaw and decompose, leading to a significant increase in $CO_2$ and methane concentrations.

The second feedback is about the ocean. The ocean is a potentially unstable reservoir of greenhouse gases. It contains at least 50 times more carbon dioxide than is found in the Earth's atmosphere. The world ocean absorbs a part of the anthropogenic $CO_2$ emissions. However, the ocean's ability to dissolve carbon dioxide decreases with temperature and may change due to Henry's law [38]. The analysis of the additional amount of $CO_2$ released into the atmosphere due to the increase in global temperature published so far for the first two scenarios is insufficient. This is because of the exceptional complexity and interdisciplinary nature of the problem. Our vision is presented in Figure 2. Below, we propose a flowchart of various scenarios of greenhouse catastrophes: the "moist" greenhouse catastrophe, the "clathrate gun" scenario, and a combined anthropogenic greenhouse catastrophe. The latter includes all positive feedback that currently exists in the Earth's climate system.

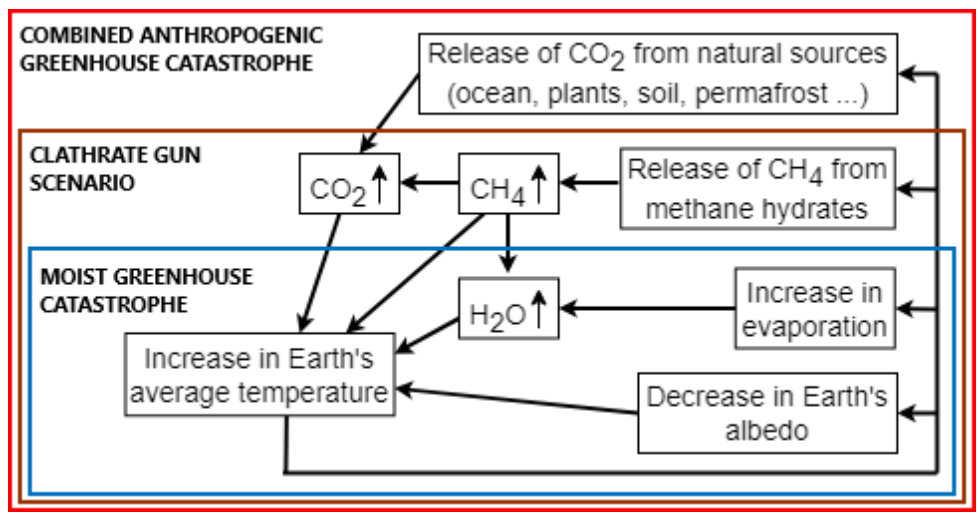

Figure 2: The flowchart of possible greenhouse catastrophes. A moist greenhouse catastrophe may occur with an increase in Earth's surface temperature triggered by water vapor in the atmosphere due to evaporation, leading to a temperature increase (positive feedback). This scenario is unlikely due to the slow increase in the Sun's luminosity. The clathrate gun scenario is possible due to the decomposition of methane hydrates and global warming. The combined anthropogenic greenhouse catastrophe suggests a set of positive feedback caused by additional $CO_2$ emissions from natural and potentially unstable carbon reservoirs [39, 40].

The goal of this work is to assess the possibility of the Runaway Greenhouse effect scenario: Greenhouse catastrophe combined with anthropogenic forcing. The first step is to estimate the ECS as accurately as possible. The second step is to introduce SIM in Climatology. In this article, we use SIM to improve the accuracy of the ECS by analyzing changes in the heat content of various components in the Earth's climate system.

## 2. System identification method (SIM)

SIM is a set of mathematical methods for studying dynamic systems that rely on observational data. The difference between SIM and traditional statistical methods of evaluation is based on the availability of additional knowledge about the dynamic system. SIM was introduced by the mathematician Gauss [41], who used the least squares method to calculate the orbital parameters of planets in the Solar System. In addition to astronomical data, he used Kepler's Second Law to calculate the angular coordinates of the planets with relative accuracy.

We propose a test before using SIM to estimate the Earth's climate parameters. A certain body with an unknown mass $m$ and specific heat capacity $C$=4.2 kJ/kg °C is supplied with energy that changes the temperature of this body ∆T(t) °C for 125 seconds. A thermos filled with water is heated by an electric device (a heater) with known power. The device power and the temperature grow exponentially. The limitation of the problem is that we can measure the temperature from the 1st to the 125th second, and the heater power from the 71st to the 118th.

The task is to determine the mass of a body (in this case, water) by measuring the temperature $\Delta T(t)$ in degrees Celsius and the heater power $W(t)$ in kilowatts per second. Mathematically, this task can be formulated as follows: let us have a discrete set of "measured experimental values" of the heater power $W$ (i, t) and the temperature of the water in the thermos $- \Delta T(i,t)$:

$$
\begin{cases}
W(i,t) = W_0 \cdot \left(e^{\frac{t}{\tau_0}} - \delta_{WS}\right) + \delta_W(i,t), & t \in [71,118]; i \in [1,20000] \\
\Delta T(i,t) = \frac{W_0}{C \cdot m} \cdot \left(\tau_0 \cdot e^{\frac{t}{\tau_0}} - \delta_{WS} \cdot t\right) + \delta_T(i,t), & t \in [1,125]; i \in [1,20000]
\end{cases}
, \quad (1)
$$

Where $i$ is the experimental number and $t$ is time (in seconds); $\delta_w(i,t)$ and $\delta_T$ (i,t) are the random variables describing the measurement errors of power and temperature, respectively; $\delta_{WS}$ is the parameter describing the heat losses; the $W_0 = 0.084$ kW and the $\tau_0 = 50$ s (time constant) are the parameters determining the change in heating power; $m = 1$ kg is the mass of the heated water.

After the data set $W(i,t)$ and $\Delta T(i,t)$ is generated based on the formula (1), the heat capacity $C = 4.2 \ kJ \ kg^{-1} \ °C^{-1}$ (of water). From this point on, the remaining parameters $W_0$, $\tau_0$ and $m$ are considered unknown. To solve the task of finding the mass $m$, we should use the expression for the increment of the heat content of the body $\Delta Q$, through the growth of its temperature $\Delta T$:

$$\Delta Q = m\, C \Delta T \qquad => \qquad m = \frac{\Delta Q}{C \Delta T} = \frac{Q(118) - Q(71)}{C\left(\Delta T(118) - \Delta T(71)\right)}. \tag{2}$$

The numerical estimate of the quantities $Q(118) - Q(71)$, $\Delta T(71)$, $\Delta T(118)$ based on the available data sets $W(i,t)$ and $\Delta T(i,t)$ based on the three methods:

The first method is a "standard statistical" method that does not use a priori information. To calculate $\Delta Q$=Q(118)- Q(71), standard integration (trapezoidal rule) is performed based on the data on the change in power *W(i,t)*. The values ΔT(71) and ΔT(118) are estimated by averaging a certain number of measurements ΔT(i,t) in the range of time *t* = 71 s and *t* = 118 s.

The second method involves using elements of SIM by analyzing the exponential growth of temperature data. In this method, the calculation of ΔQ=Q(118)-Q (71) is done by estimating the exponential regression coefficients for the set W(i,t). To determine the values of ΔT(71) and ΔT(118) in the second method, exponential regression coefficients for the set *ΔT(i,t)* are also calculated, followed by calculating the regression function at *t*=71 s and *t*=118 s.

The third method involves a SIM that utilizes the linearity property of the system under study. This is evident in the equal time constant of the exponential growth of heater power and temperature in such a system. The calculation of ΔQ = Q(118)- Q (71) follows a similar approach to the second method, with the key difference being that the time constant for estimating the regression coefficients for ΔT(i,t) is assumed to be equal to the time constant for W(i,t). Due to the significantly lower relative error in the "measurements" of W(i,t) compared to ΔT(i,t), this results in a substantial reduction in both random and systematic errors in determining m.

The results of estimating the value of mass *m* using the three methods described above are presented in Fig. 3. The improved SIM provides the most accurate solution to the task.

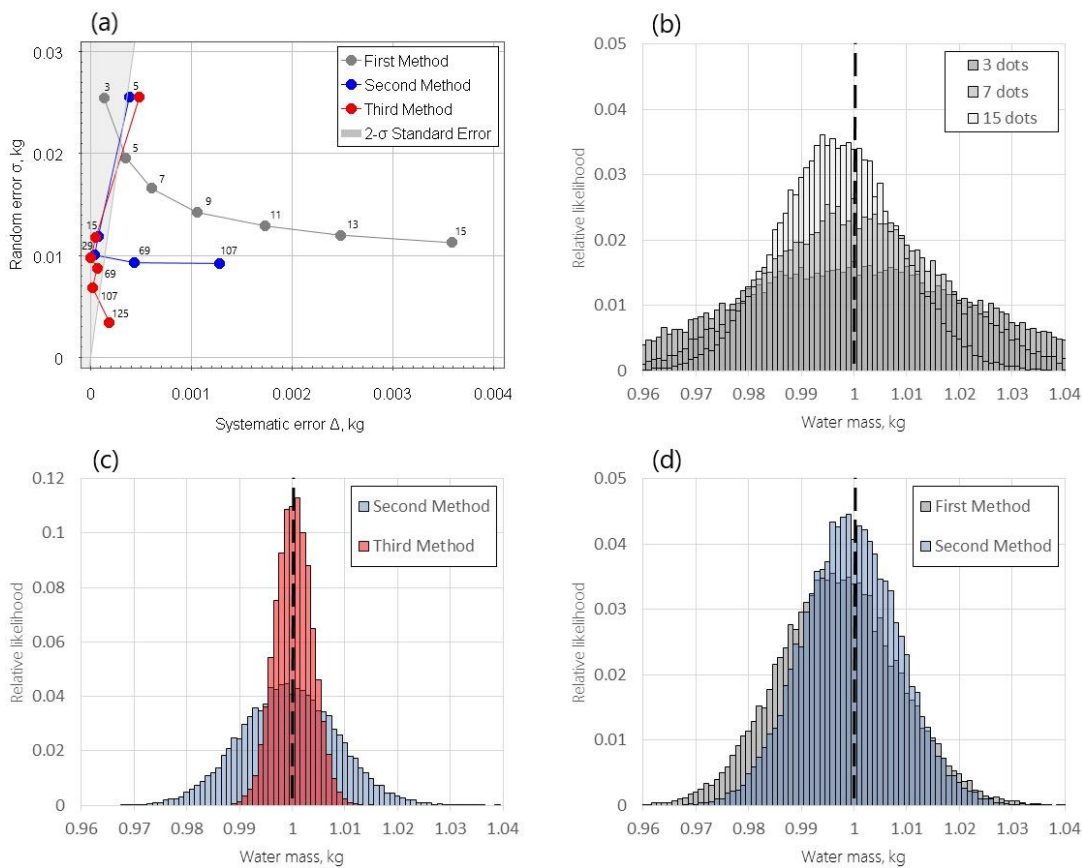

Figure 3: The results of solving the test task of finding the mass of heated water using three different estimation methods. Diagram (a) displays the dependencies of random and systematic errors in determining the mass for all three methods based on the number of measurements for averaging (first method) or determining regression coefficients (second and third methods). The improved SIM method (third) shows a significantly lower level of random and systematic errors. Histograms (b) illustrate the limitations of the standard statistical method: while the random error component decreases with an increase in the number of time points for averaging, the systematic error increases. Histograms (c) and (d) highlight the advantage of the improved SIM over the standard (c) and the standard SIM over the standard statistical method (d), respectively.

The analogy between the test task of determining the mass of heated water in a thermos and the problem of determining the equilibrium climatic sensitivity is illustrated in Fig. 4. Firstly, we observe the exponential growth of $CO_2$ concentration in the atmosphere and the average planetary temperature of Earth (Fig. 4). It is also worth noting that all the model curves depicting exponential growth in Fig. 4 share the same time constant $\tau_0 = 47.35 \pm 0.16$ years. This is because the $CO_2$ concentration is measured with significantly less random error compared to the measurement error of the average planetary temperature. This forms the basis for using the SIM, like the third method used to evaluate the equilibrium climatic sensitivity (ECS).

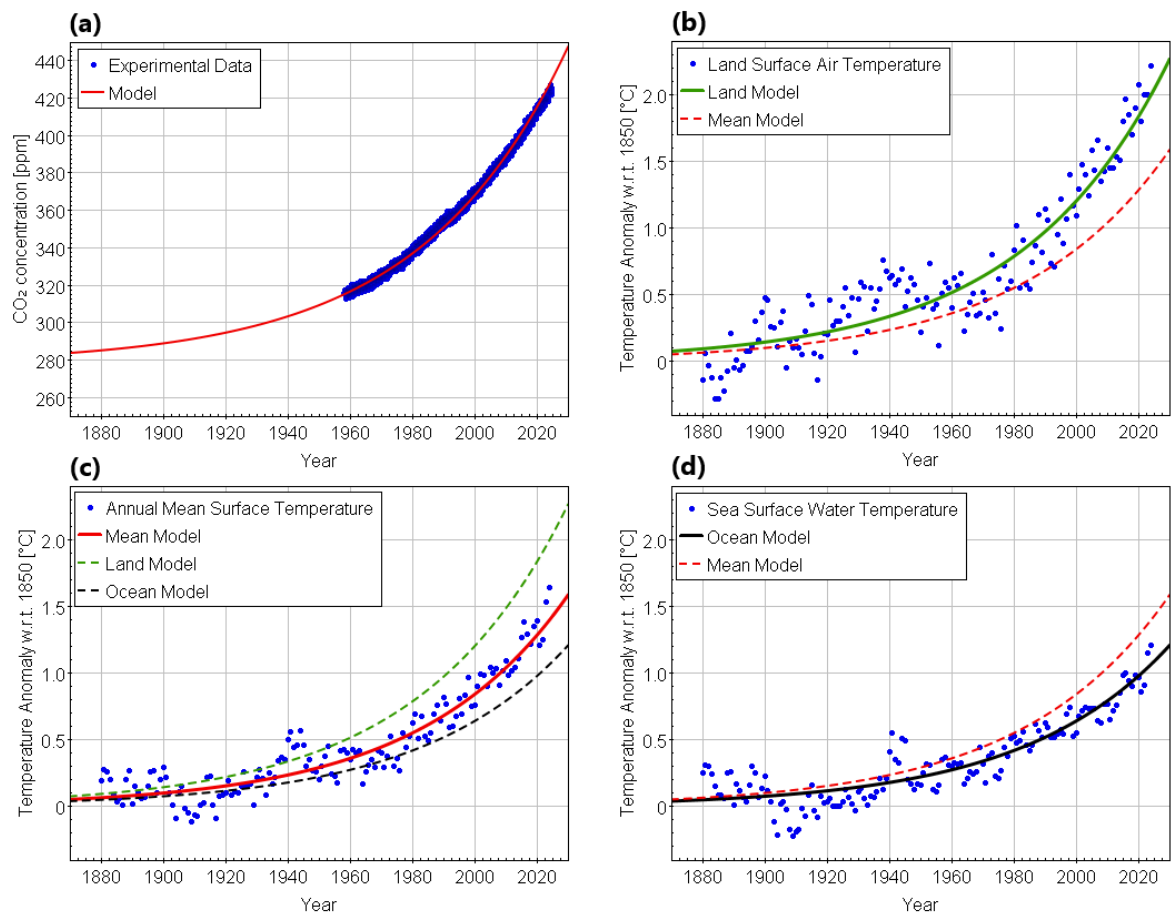

**Figure 4.** Graphs of the exponential growth. (a) $CO_2$ concentration in the atmosphere. Over 200 years of industrial development, the concentration of carbon dioxide in the atmosphere has risen from 280 to 420 ppm with annual growth 2-3 ppm·yr$^{-1}$ [42]. (b) Land surface air temperature (LSAT) anomaly [42]. (c) Annual mean global surface air temperature (GSAT) anomaly. Over 200 years of industrial development, GSAT anomaly has risen from 0 to 1.6 °C with annual growth 0.03 °C·yr$^{-1}$ [42]. (d) Sea surface water temperature (SST) anomaly [42]. It is evident that the data for the $CO_2$ concentration in the atmosphere have the smallest scatter in the graph (a). The red curves in all graphs are exponential approximations of the measured data with a characteristic time $\tau_0 = 47.35 \pm 0.16$ years, which was obtained from the data in the graph (a).

We assert the estimates of climate parameters given in the 6$^{th}$ IPCC assessment report to eliminate the ambiguity of the results of the ECS assessment (Table 1).

**Table 1.** Summary table of parameters. $\Delta F$ is the total anthropogenic effective radiative forcing for 1750-2019 ([43], p. 960). $\Delta Q$ (total) is the global energy inventory for 1971-2018 ([43], p. 938). $\Delta Q(O)$, $\Delta Q(L)$, $\Delta Q(C)$, $\Delta Q(A)$ are the energy inventory for 1971-2018 of the ocean, land, cryosphere and atmosphere respectively ([43], p. 938). $\Delta F(2xCO_2)$ is the effective radiative forcing to $2 \times CO_2$ change since pre-industrial times ([43], p. 945). 1 ZJ = $10^{21}$ J.

| Variable: | $\Delta F$ | $\Delta Q(total)$ | $\Delta Q(O)$ | $\Delta Q(L)$ | $\Delta Q(C)$ | $\Delta Q(A)$ | $\Delta F(2xCO_2)$ |
|---|---|---|---|---|---|---|---|
| Units: | W m$^{-2}$ | ZJ | ZJ | ZJ | ZJ | ZJ | W m$^{-2}$ |
| 5-95% uncertainty range | [1.96 to 3.48] | [324.5 to 545.3] | [285.7 to 506.2] | [18.6 to 25.0] | [9.0 to 14.0] | [4.6 to 6.7] | [3.46 to 4.40] |

The flowchart of the general Monte-Carlo method used to calculate the confidence intervals of the numerical ECS estimates is presented in Fig. 5.

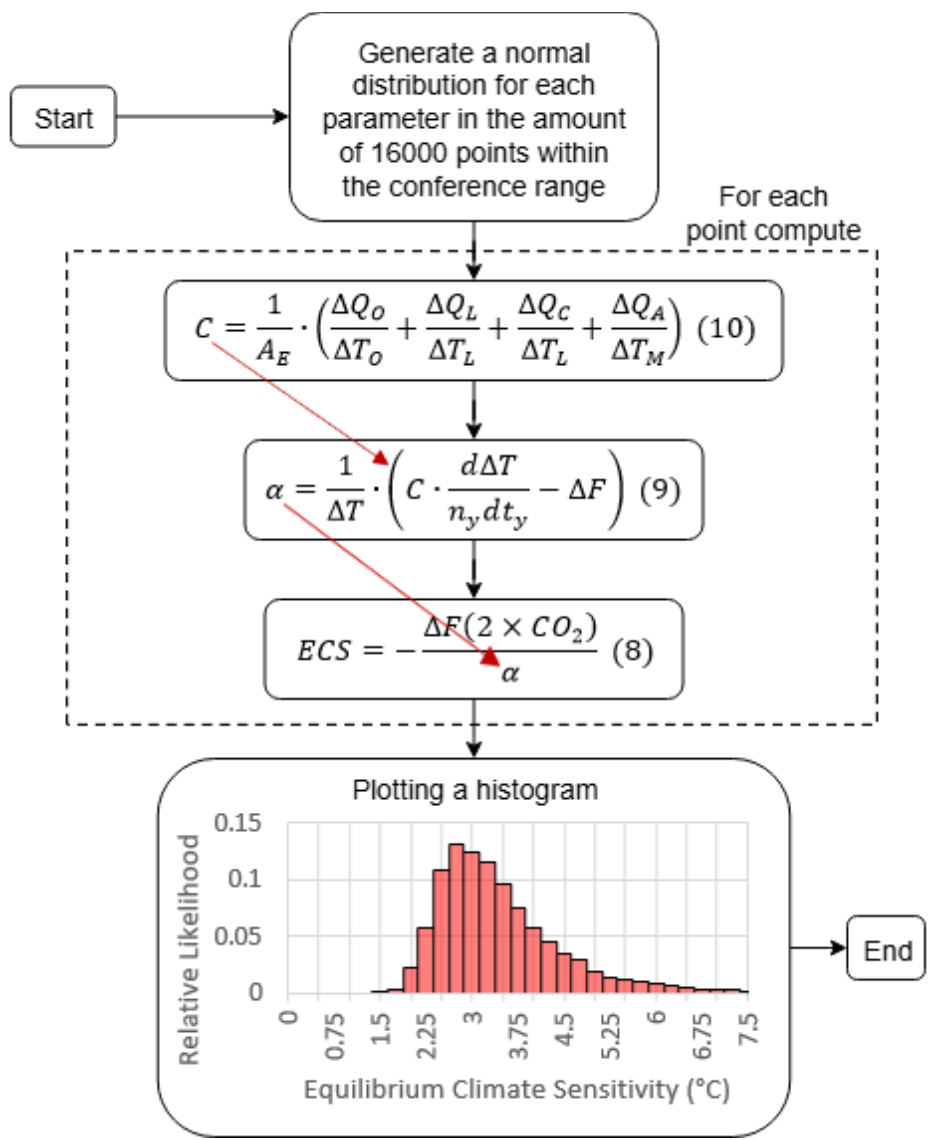

**Figure 5:** The flowchart depicts the Monte-Carlo method based on a normal distribution with 16,000 points for each parameter within the confidence range. Subsequently, each of the 16,000 points for each parameter is substituted into formulas to obtain the distributions for heat capacity (C), feedback parameter (α), and equilibrium climate sensitivity (ECS). The confidence range of 5-95% is then determined for each value by calculating the 5th and 95th percentiles.

## 3. Results.

The Global Surface Temperature (GST) response to perturbations related to energy imbalance is approximated by the linear energy budget equation, in which $\Delta N$ represents the change in the TOA net energy flux, $\Delta F$ is an Effective Radiative Forcing (ERF) perturbation to the TOA net energy flux, $\alpha$ is the net feedback parameter, and $\Delta T$ is the change in GST:

$$\Delta N = \Delta F + \alpha \Delta T \qquad (3)$$

The ERF ($\Delta F$ units: W m$^{-2}$) quantifies the change in the net TOA energy flux of the Earth system due to an imposed perturbation (e.g., changes in greenhouse gas or aerosol concentrations, incoming solar radiation, or land-use change). The $\Delta F$ value can be divided into components associated with different sources, such as different greenhouse gases.

$$\Delta F = W(CO_2) + W(CH_4) + W(N_2O). \qquad (4)$$

There is no term in Eq. 4 associated with changes in the luminosity of the Sun. Only small quasi-periodic changes in luminosity have been measured at a level of 0.1% over an interval of about a year and 0.5% within the 11-year cycle of solar activity, as of today. The evolution of the Sun as a star predicts an increase in its luminosity by 1% every 110 million years. This is negligibly small and has not yet been confirmed by direct measurements of the Sun's luminosity. The feedback parameter $\alpha$ [W m$^{-2}$ °C$^{-1}$] quantifies the change in the energy flux at TOA for a given change in GST:

$$\alpha = \alpha(Plank) + \alpha(water\ vapor) + \alpha(albedo) \qquad (5)$$

The $\Delta N$ is the energy imbalance expressed as the derivative of the change in global energy inventory ($\Delta Q$) with respect to time.

$$\Delta N = \frac{d(\Delta Q)}{dt} = \frac{d(C\Delta T)}{dt} = C \cdot \frac{d\Delta T}{d(n_y t_y)} = C \cdot \frac{d\Delta T}{n_y dt_y}, \qquad (6)$$

where $C$ is the total planetary heat capacity of Earth's surface, including the atmosphere, the ocean, and the landmass, $t$ is time in years ($3.16 \times 10^7$ s). We can rewrite Equation (3) by considering Equation (6):

$$C \cdot \frac{d\Delta T}{n_y dt_y} = \Delta F + \alpha \Delta T. \qquad (7)$$

Equilibrium climate sensitivity (ECS) is calculated using Eq (7) if the concentration of $CO_2$ in the atmosphere doubles.

$$ECS = \Delta T(2 \times CO_2) = -\frac{\Delta F(2 \times CO_2)}{\alpha}. \qquad (8)$$

The magnitude of the change in radiative forcing $\Delta F(2 \times CO_2)$ with doubling of the concentration of $CO_2$ was estimated based on numerical modeling of the atmosphere. There are several ways to estimate the $\alpha$ coefficient. Direct estimates are based on numerical 3D models of the atmosphere change using instrumental data (Eq. 7) including the heat content of the Earth's surface, and the ocean.

$$\alpha = \frac{1}{\Delta T} \cdot \left( C \cdot \frac{d\Delta T}{n_y dt_y} - \Delta F \right). \qquad (9)$$

We define the total heat capacity as the sum of the components from the ocean, land, cryosphere, and atmosphere:

$$C = \frac{1}{A_E} \cdot \left( \frac{\Delta Q_O}{\Delta T_O} + \frac{\Delta Q_L}{\Delta T_L} + \frac{\Delta Q_C}{\Delta T_L} + \frac{\Delta Q_A}{\Delta T_M} \right) = [10.2 \ to \ 17.9] \ 10^8 \ J \ m^{-2} \ °C^{-1}, \qquad (10)$$

where $A_E = 5.1 \ 10^{14}$ m$^2$ is the surface area of the Earth; $\Delta Q_O$, $\Delta Q_L$, $\Delta Q_C$, $\Delta Q_A$ are the energy inventory for 1971-2018 of the ocean, land, cryosphere and atmosphere respectively; $\Delta T_O$, $\Delta T_L$, $\Delta T_M$ are the change in temperature over the ocean, temperature over land and average temperature for the same years, respectively. Energy inventory values are taken from Table 1 and the values of temperature changes are taken from our temperature regressions according to NASA/GISS/GISTEMP data [42] (Fig.4):

$$\Delta T_M = [0.80 \ to \ 0.88] \cdot \exp \left( \frac{t - 2000}{[47.1 \ to \ 47.6]} \right), \qquad (11)$$

$$\Delta T_L = [1.13 \ to \ 1.26] \cdot \exp \left( \frac{t - 2000}{[47.1 \ to \ 47.6]} \right), \qquad (12)$$

$$\Delta T_O = [0.60 \ to \ 0.68] \cdot \exp \left( \frac{t - 2000}{[47.1 \ to \ 47.6]} \right). \qquad (13)$$

The coefficient $\alpha$ is equal to $[-1.91 \ to - 0.56] \ W \ m^{-2} \ °C^{-1}$. It is comparable with the value obtained by IPCC6 ([43], p. 978) using a 3D model of the atmosphere $\alpha(IPCC6) = [-1.81 \ to - 0.51] \ W \ m^{-2} \ °C^{-1}$. Our estimate of $ECS(this \ work) = [2.0 \ to \ 7.0]$ °C differs slightly from the 3D atmospheric models provided in the IPCC6 report ([43], p. 994):

$ECS(IPCC6) = [2.1 \; to \; 7.7] \; ^{\circ}C.$

The ECS histograms were built using the Monte-Carlo method (Figure 6).

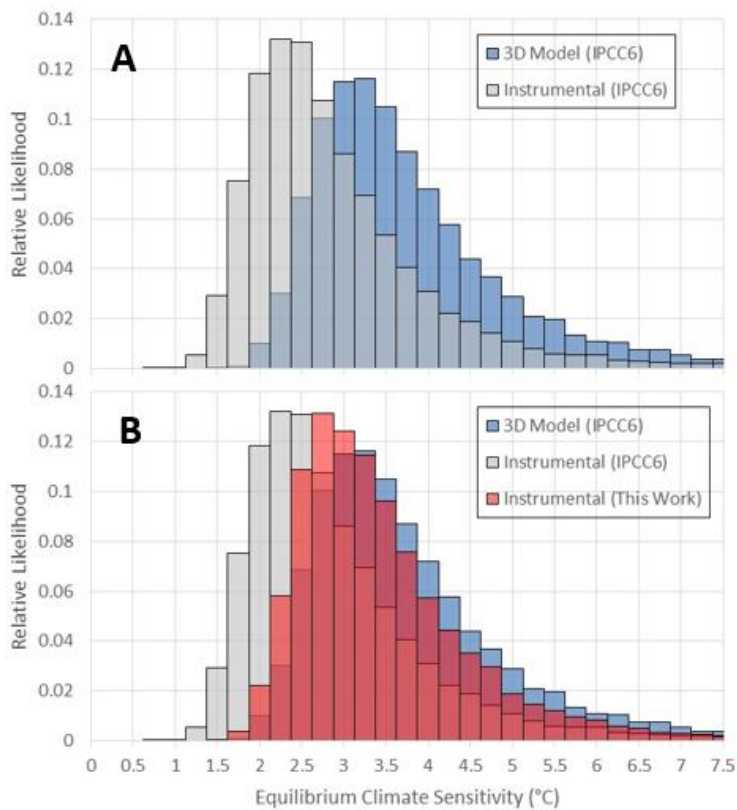

Figure 6: Histograms of Equilibrium Climate Sensitivity (ECS) estimations. The blue histogram is constructed using IPCC6 data ([43], p. 994) derived from 3D modeling methods (A and B). The gray histogram corresponds to the estimates given in IPCC6 ([43], p. 996) and derived from an analysis of instrumental heat inventory data for the main elements of the climate system. The red histogram uses the same instrumental heat inventory data for the elements of the climate system but employs the improved system identification method (SIM) described in this paper. All histograms are obtained using the Monte-Carlo method. The detailed process of constructing the red histogram is available as an online animation.

## 4. Summary.

The System Identification Method (SIM) used in this study allows for a more accurate estimation of Equilibrium Climate Sensitivity (ECS) (Fig. 6B, red histogram). SIM can also potentially be applied to refine the ECS using available data for: (i) estimating the different rates of land and ocean temperatures; (ii) assessing paleoclimatic changes based on the analysis of the gradual disintegration of ice shields in Antarctica and Greenland. SIM could potentially improve the accuracy of ECS estimation using hybrid methods, including 3D modeling.

The planetary heat capacity should be evaluated using the contributions of the ocean, land, cryosphere and atmosphere.

The analysis indicates a shift of the ECS upward instead of significantly low estimates of the ECS based on past instrumental data analysis (Fig. 6A, gray histogram). The ECS value increases significantly as the Earth's global surface temperature (GST) grows, and the spectral range of the transparency windows in Earth's atmosphere narrows [44].

Highly accurate evaluation of the ECS is important to assess the possibility of the Runaway Greenhouse effect scenario.